\documentclass{aa}
\def\mdens{{\rm g~cm^{-3}}}

\def\bdens{{\rm fm^{-3}}}
\def\msun{{\rm M}_\odot}
\def\rhonuc{\rho_{_0}}
\def\nbnuc{n_{_0}}
\def\rhored{\overline{\rho}}
\def\nbred{\overline{n}}
\def\rnuc{r_{_0}}

\usepackage{amssymb,amsmath}

\usepackage{graphicx}
\usepackage{color}

\usepackage{bm}
\begin{document}

\title{Neutron stars with hyperon cores: stellar radii and equation of state near
nuclear density}
\author{ M.~Fortin \and
 J. L. Zdunik \and P. Haensel \and M. Bejger}
\institute{N. Copernicus Astronomical Center, Polish
           Academy of Sciences, Bartycka 18, PL-00-716 Warszawa, Poland
\\
{\tt fortin@camk.edu.pl, jlz@camk.edu.pl, haensel@camk.edu.pl, bejger@camk.edu.pl}}
\offprints{J.L. Zdunik}
\date{Received 13 August 2014 Accepted 23 January 2015}
\abstract{{The existence of $2~\msun$ pulsars puts very strong
constraints on the equation of state (EOS) of neutron stars (NSs)
with hyperon cores, which can  be satisfied only by  special models
of hadronic matter. The radius-mass relation for these models is
sufficiently specific that it could be subjected to an observational test with
future X-ray observatories.}}
{We want to study the impact of the presence of hyperon cores on the radius-mass
relation for NS. We aim to find out how, and for which particular stellar mass
range, a specific relation $R(M)$, where $M$ is the gravitational mass, and $R$ is
the circumferential  radius, is associated with  the presence of a hyperon core.}
{We consider  a  set of 14 theoretical EOS of
dense matter, based on the relativistic mean-field (RMF)
approximation, allowing for the presence of hyperons in NSs.
We also discuss a recent EOS
based on non-relativistic G-matrix theory
yielding NSs with hyperonic cores and $M>2\;\msun$. We seek 
correlations between $R(M)$ and the stiffness of the EOS below the
hyperon threshold needed to pass the $2~\msun$ test.}
 {For NS masses  $1.0<M/\msun<1.6$, we get  $R>13~$km,
because of a very stiff  pre-hyperon segment of the EOS. At
nuclear density ($n_{_0}=0.16~~\bdens$), the pressure is
significantly higher than  a robust upper bound obtained recently 
using chiral effective field theory.}
{If massive NSs  do have a sizable hyperon core,
then according to current models the radii for
$M=1.0-1.6~\msun$ are necessarily $>13 ~$km.  If, on the contrary,
 a NS with a radius $R^{\rm (obs)}<12~$km  is
observed in this mass domain, then sizable hyperon cores in NSs, as we model them now, are
 ruled out. Future X-ray missions with $<5\%$
precision for a simultaneous $M$ and $R$ measurement will have the
potential to solve the problem with observations of NSs.
Irrespective of this observational test, present EOS  allowing for hyperons that
fulfill condition $M_{\rm max}>2~\msun$ yield a pressure at nuclear density
that is too high relative to up-to-date microscopic calculations
of this quantity.}

\titlerunning{Hyperon cores and neutron star radii}
\authorrunning{}
\maketitle

\section{Introduction}
\label{sect:introduction}
Hyperons, which are baryons containing at least one strange quark, 
were discovered more than 50 years ago. They are frequently studied in terrestrial
laboratories. Although unstable on Earth, it is expected however
that they are stably present in the dense interiors of neutrons
stars (NSs). Recent measurements of $2\;\msun$ pulsars \citep{Demorest2010,Antoniadis2013}
represent a
challenge for equations of state (EOS) of NSs with hyperon cores. The
difficulty of reaching a high mass is related to a significant
softening of the EOS associated with the hyperonization of the
matter.  With one exception, only EOS based on the relativistic
mean-field (RMF) models, after tuning of their Lagrangians, turned
out to be able to produce NS with maximum allowable mass $M_{\rm
max}>2\;\msun$ and sizable hyperon cores (see \citealt{Colucci2013}
and references therein). {A very recent paper of \citet{Yamamoto2014}
is based on  the non-relativistic G-matrix theory of dense baryon
matter. Their MPa model yields $M_{\rm max}>2\;\msun$ for NSs with
hyperon cores. As we will see, however, the properties of NS models
for  MPa EOS are consistent with those obtained by our set of 
RMF EOS.
These successful models with hyperonization of  NS
matter (one of them being hereafter referred to as EOS.H) merit a  careful inspection.
We will focus here on two specific features of these models: NS radii and
 EOS at pre-hyperon density; as we will show, both features are
interrelated.

The EOS  below nuclear density $\rho_{_0}=2.7\times 10^{14}\;\mdens$
(corresponding to the baryon number density $n_{_0}=0.16~\bdens$) is
commonly believed to be rather well known (see \citealt{Hebeler2013}).
However, to construct a complete family of NS models, up to the
maximum allowable mass $M_{\rm max}$, one needs the EOS for up to
$\sim 8\rhonuc$. Nuclear densities $\rhonuc$ and $\nbnuc$ are
suitable units for the mass-energy density $\rho$ and the
baryon number density $n_{\rm b}$ in the NS core. In what follows we will
use dimensionless (reduced) densities $\rhored\equiv\rho/\rhonuc$
and $\nbred\equiv n_{\rm b}/\nbnuc$. Because of uncertainties in the
theory of dense matter, the only chance to unveil the actual EOS of
 degenerate matter at supra-nuclear density relies on the
observations of NSs.

Mathematically, both $M_{\rm max}$ and the radius of NS of
(gravitational) mass $M$,  $R_{_M}\equiv R(M)$,  are functionals of
the EOS. We expect that NS matter at $0.5<\rhored<2$ (the so-called
outer core)  is composed mostly of neutrons, with a few percent
admixture of protons, electrons, and muons. At higher densities,
hyperons or even quark gluon plasma might appear, forming a
strangeness carrying NS core. In the present paper, we restrict
ourselves to the NSs cores where quarks are confined into baryons:
nucleons and hyperons.
The EOS fulfilling $M_{\rm max}>2\;\msun$ and
allowing for the presence of hyperons in NSs cores form the set {\bf
EOS.H}.

Independent of the uncertainties related to the structure of  the
cores of NSs, any theoretical EOS.H has to be consistent with the
semi-empirical parameters of nuclear matter at $\rhored\approx 1$.
Moreover, the hyperon component should be consistent with
semi-empirical estimates (potential wells,
$\Lambda\Lambda$-interaction) coming from hypernuclear physics \citep{SB08}.
Basic features of {\bf EOS.H} resulting from various theoretical
models and puzzles, which are as yet not resolved, are briefly
summarized in Sect.\;\ref{sect:EOS.overv}. Another constraint,
related to the value of the pressure $P$ at nuclear (saturation)
density, is discussed in Sect.\;\ref{sect:EOS.N.H}.

The impact of the uncertainty in the EOS on the mass vs. central
density dependence for NS is described in
Sect.\;\ref{sect:Mnc.Mrhoc}. We  prefer to use
$\nbred_{\rm c}$ instead of $\rhored_{\rm c}$ because the
former  characterizes the degree of packing of baryons at the
NS centre. Different segments (domains) of the EOS
$P(\nbred)$ determine measurable global stellar parameters in
different NS mass domains. The radius of a $1.4\;\msun$ NS,
$R_{1.4}$, is mostly determined by EOS$(1<\nbred<3)$, while the
value of $M_{\rm max}$ is to a large extent determined by
EOS$(4<\nbred<7)$.

 As we show in Sect.\; \ref{sect:RM.hyp}, the NS radius for
 $1.0\;\msun<M<1.6\;\msun$  for {\bf EOS.H} is larger than 13 km.
This seems to be  an unavoidable  consequence of $M_{\rm
max}>2.0\;\msun$ condition,
  which  implies a very high stiffness of the pre-hyperon
  (i.e., purely nucleon)  segment $1<\nbred\lesssim 2 - 3$
  of {\bf EOS.H}.

Sect.\;\ref{sect:MPc} explores the difference in the values of NS
masses at a given central pressure and connects it with the difference
in pressure distribution within NS models. In
Sect.\;\ref{sect:RM.causal} the causal-limit EOS is used to provide
a bound on the NS radius, $R(M)$. Sect.\;\ref{sect:RM.rot} is
devoted to the effects of  NS rotation.

  In Sect.\;\ref{sect:RM.obs} we discuss  a possible  meaning of large radii
  of NSs  with hyperon cores    in the context of recent
  measurements of radii of NSs.    Our  conclusions are formulated in
   Sect.\;\ref{sect:conclusions}.

Preliminary results of our work were presented at  the Nuclear
Physics in Astrophysics VI conference, Lisbon, Portugal, May 19-24,
2013, at  the EWASS 2013 Symposium "Extreme physics of neutron stars" at
Turku, Finland, July 10-13, 2013, at  EMMI Meeting, FIAS, Frankfurt,
Germany, October 7-10, 2013,  and at the conference, "The structure
and signals of neutron stars, from birth to death", Florence, Italy,
March 24-28, 2014.
\section{Equation of state  of neutron-star matter}
\label{sect:EOS.overv}
Except for a very outer layer, whose contribution to NS mass and
radius can be neglected, the matter in NS interior is strongly
degenerate, and can be approximated by that calculated at $T=0,$
assuming the ground state composition (cold catalyzed matter, see
e.g. \citealt{NSbook2007}). For these EOS, pressure depends on the
density only.

In general theory of relativity (GTR) the matter density is defined
as $\rho={\cal E}/c^2$, where ${\cal E}$ is the total energy density
 (including rest energy of
particles). Baryon density $n_{\rm b}$ is defined as the baryon number
(baryon charge) in a unit volume. Using elementary thermodynamics
one obtains  the  relation
 between  $P$  and $\rho$ from  the calculated function
  ${\cal E}(n_{\rm b})$,
\begin{equation}
P(n_{\rm b})=n_{\rm b}^2 {{\rm d}({\cal E}/n_{\rm b})\over {{\rm d}n_{\rm b}}}~,
 ~~~\rho(n_{\rm b})={\cal E}(n_{\rm b})/c^2 ~\Longrightarrow P=P(\rho)~.
 \label{eq:P.nb.rho}
\end{equation}
\subsection{EOS satisfying the semi-empirical nuclear-hypernuclear constraints
and $M_{\rm max}>2.0\;\msun$}
\label{sect:EOS.B}

The models  of the {\bf EOS.H} set reproduce (within some tolerance)
four semi-empirical values of parameters of nuclear matter at
saturation: saturation density $n_{\rm s}$, energy per nucleon
$E_{\rm s}$, symmetry energy $S_{\rm s}$ and incompressibility
$K_{\rm s}$. The semi-empirical value of a fifth parameter, the
density slope of symmetry energy $L_{\rm s}$, is relatively poorly
known \citep{Hebeler2013} and hence it is not imposed as a
constraint. Together with $M_{\rm max}>2\;\msun$, this amounts to  five
constraints imposed on an EOS belonging to  {\bf EOS.H}.

 Apart  from the  five constraints described above, a given EOS.H has to
reproduce additional semi-empirical values of three potential
wells of (zero momentum) hyperons in nuclear matter at $n_{\rm s}$:
these include the sixth, seventh, and eighth constraints on {\bf EOS.H}.
Moreover, for most of the EOS.H models, a ninth condition, 
fitting a semi-empirical estimate of the depth of potential well of
$\Lambda$ in $\Lambda$-matter, is also imposed.  In summary, there
are eight or nine constraints to be satisfied by an EOS.H.
Constructing an EOS of the {\bf EOS.H} set is therefore associated
with a strong tuning of the dense matter models. With the exception of a
very recent EOS.H of \citet{Yamamoto2014}, only the EOS based on the
non-linear RMF theories can satisfy these
conditions.

Selecting a very specific type of approximation, the RMF ,  is
a {\it \textup{first tuning}} of the dense matter model. Moreover, 
to satisfy $M_{\rm max}>2\;\msun$, baryon fields are not only coupled  to (standard) $\sigma$, $\omega$, and $\rho$ meson fields.
Namely, the hyperon fields are additionally coupled to the vector
$\phi$ meson field. The addition of $\phi$ (and possibly also of a
$\sigma^\star$ meson field, which provides a scalar coupling between
hyperons only) is a {\it \textup{second tuning}}. All EOS from {\bf EOS.H}
are able to satisfy $M_{\rm max}>2.0\;\msun$ due to repulsion
produced by the $\phi$-meson coupled only to hyperons. Moreover, in
some cases, an amplification of the hyperon repulsion due to
SU(6)-symmetry breaking in the vector-meson coupling to hyperons is
introduced, which is   a {\it \textup{third tuning}}.
 Finally, some models have density-dependent coupling constants of
baryons to meson fields. This allows for a {\it \textup{fourth tuning}}. The
models included in {\bf EOS.H} set are listed in Table 1 together
with brief characteristics and references.

Among the eight models consistent with a $2\;\msun$ NS in
\citet{Sulaksono2012}, we selected four of them, three with SU(6)
symmetry, the stiffest, the softest and an intermediate one, and one
with SU(6) symmetry broken. \citet{LM14} obtain similar results to
\citet{Weissenborn2011phi} and their EOS are not included in  Table
1. We also restrict ourselves to EOS calculated for a zero
temperature. The EOS at finite temperature, such that they are suitable
for supernova and proto-neutron star modelling and including a
transition to hyperonic matter, have also been developed. In
particular \citet{BH14} using a RMF model obtain a maximum mass for
cold NS of $M_{\rm max}=2.1\;\msun$. \citet{GR13}, extending the EOS
by \citet{LS} to include hyperons with the model by \citet{BG}, reach
$M_{\rm max}=2.04\;\msun$; however, they include only the $\Lambda$
hyperon.

In order to have a reference set of models of  widely used purely
nucleon EOS, denoted as ${\bf EOS.N_{\rm\bf ref}}$, we selected
three models listed in the upper part of Table 1. These EOS
produce "standard NS", are consistent with all semi-empirical
constraints and yield $M_{\rm max}>2.0\;\msun$. We note
that not only is $L_{\rm s}$ for {\bf EOS.H}  widely scattered ($62 - 118$
MeV) but it is also typically much higher than for ${\bf EOS.N_{\rm\bf ref}}$
($37-59$ MeV).

The EOS of the core is supplemented with an EOS of NS crust. We
assume that the crust is composed of cold catalyzed matter. For the
very outer layer with $\rho<10^8~\mdens$ we use classical BPS EOS
\citep{BPS1971}. The outer crust with $\rho>10^8~\mdens$ is
described by the EOS of \citet{HP1994}, while for the inner crust we
apply the SLy EOS of \citet{DH2001}. A smooth matching with an
interpolation between the crust and core EOS is applied to get a
complete EOS of NS interior.
\footnote{ The non-uniqueness of the
crust-core matching in the EOS introduces some indeterminacy, which
only disappears  for unified EOS where the crust and core EOS are
based on the same nuclear many-body model, starting from  the same nucleon
interaction (see e.g. \citealt{DH2001,GP2014}).}

\parindent 21pt

\begin{table*}
\begin{center}
\caption{Equations of state. For ${\bf EOS.N_{\rm\bf ref}}$ (upper
part) we selected  three widely used EOS, which produce standard
values of NS parameters. The lower part of the table contains our {\bf
EOS.H} set. For further explanation, see the text.}
\begin{tabular}[t]{|c|c|c|}
\hline\hline
&&\\
 EOS & theory  & reference \\
&&\\
\hline
\hline
&&\\
 APR    & Variational, infinite chain summations  &  \citet{APR1998} \\&&\\
 DH     & energy-density  functional, Skyrme type &  \citet{DH2001}    \\&&\\
BSk20   & energy-density  functional, Skyrme type &  \citet{Fantina2013}    \\&&\\
\hline\hline &&\\
BM165   & RMF, constant couplings, SU(6)          &  \citet{Bednarek2012}\\ &&\\
DS08    & RMF, constant couplings, SU(6)          &  \citet{Dexheimer2008}\\&&\\
GM1Z0   & RMF, constant couplings, SU(6) broken   &  \citet{Weissenborn2011phi}  \\&&\\
M.CQMCC & RMF, constant couplings, SU(3)          &  \citet{Miyatsu2013} \\&&\\
SA.BSR2 & RMF, constant couplings, SU(6)          &  \citet{Sulaksono2012} \\&&\\
SA.TM1  & RMF, constant couplings, SU(6) broken   &  \citet{Sulaksono2012}  \\&&\\
G.TM1   & RMF, constant couplings, SU(6) broken   &  \citet{Gusakov2014}  \\&&\\
M.TM1C  & RMF, constant couplings, SU(3)          &  \citet{Miyatsu2013}   \\&&\\
SA.NL3  & RMF, constant couplings, SU(6)          &  \citet{Sulaksono2012}  \\&&\\
M.NL3B  & RMF, constant couplings, SU(6)          &  \citet{Miyatsu2013}    \\&&\\
M.GM1C  & RMF, constant couplings, SU(3)          &  \citet{Miyatsu2013}    \\&&\\
SA.GM1  & RMF, constant couplings, SU(6)          &  \citet{Sulaksono2012}  \\&&\\
UU1     & RMF, density-dependent couplings, SU(6) &  \citet{UU2009}  \\&&\\
UU2     & RMF, density-dependent couplings, SU(6) &  \citet{UU2009}   \\&&\\
\hline\hline
\end{tabular}
\label{tab:EOS}
\end{center}
\end{table*}
\begin{figure}[h]
\resizebox{\columnwidth}{!}
{\includegraphics[angle=0,clip]{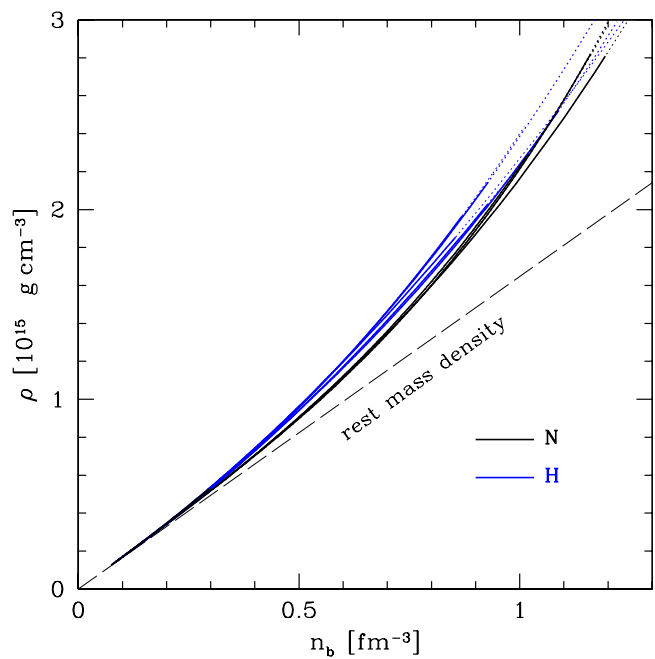}} \caption{(Colour
online) Mass density $\rho={\cal E}/c^2$ vs. baryon density $n_{\rm
b}$ for NS matter for the set of EOS presented in Figs. \ref{fig:Mnc-eos}-\ref{fig:MR}.
Dotted segments correspond to the central
densities of NS models, which are unstable with respect to radial
oscillations.
 Relation $\rho=\rho(n_{\rm b})$ deviates from
linearity for $n_{\rm b}>0.3\;\bdens$. Non-linearity grows with
increasing $n_{\rm b}$  and is EOS-dependent. For $n_{\rm b}\lesssim
0.2\;\bdens$, the linear approximation $\rho \approx n_{\rm b} m_n$
(where $m_n$ is neutron mass) is valid.}
 \label{fig:rho-nb-eos}
\end{figure}
\section{Overpressure of EOS.H at nuclear density}
\label{sect:EOS.N.H}
We start our  comparative study of {\bf EOS.H} and  ${\bf
EOS.N_{\rm\bf ref}}$  by calculating the pressure at nuclear
density,  $P^{\rm (N)}(n_{_0})$ and $P^{\rm (H)}(n_{_0}),$
respectively. Results are collected in the Appendix
Table\;\ref{tab:P0.R.RCL}. In the following $P_{33}$ refers to the
pressure $P$ in the units of $10^{33}$ dyn cm$^{-2}$. We notice a
striking difference between {\bf EOS.H} and  ${\bf EOS.N_{\rm\bf
ref}}$. The values of $P^{\rm (N)}_{33}(n_{_0})$ are concentrated
within $3.3\pm 0.3$, while the values of $P^{\rm (H)}_{33}(n_{_0})$
are significantly higher, within   $8\pm 2.5$.

\citet{Hebeler2013} argue that the  EOS of pure neutron matter at
$n_{\rm b}\le \nbnuc$ can be reliably calculated using the up-to-date
many-body theory of nuclear matter. Their results are in remarkable agreement
with those by \citet{Gandolfi2012} using an approach completely different
from that adopted by \citet{Hebeler2013}. At this density, NS matter in
beta equilibrium is expected to be somewhat softer than the pure
neutron matter. \citet{Hebeler2013} calculate the effect of the
presence of an admixture of protons and electrons in beta
equilibrium on the EOS, combining the EOS of neutron matter and
available semi-empirical information about nuclear symmetry energy
and its density dependence (slope parameter $L_{\rm s}$).
Interpolating between the values in their Table 5, we conclude that
\citet{Hebeler2013} provide the following constraint on the pressure of
NS matter at $\nbnuc$:
\begin{equation}
2.7<P_{33}(\nbnuc)<4.4\;.
\label{eq:Hebeler.P}
\end{equation}
 This constraint is satisfied by ${\bf EOS.N_{\rm\bf ref}}$. On the contrary, it
is badly violated by EOS from  {\bf EOS.H}, which give
$P_{33}(\nbnuc)$ significantly higher than the upper bound in
Eq.\;(\ref{eq:Hebeler.P}). Before considering consequences of the
''overpressure'' of the nucleon (pre-hyperon) segment of {\bf EOS.H}
for NS radii, we discuss two different parametrizations  of NS
models.

\section{Two densities and two parametrizations of neutron star models}
\label{sect:Mnc.Mrhoc}
When investigating the EOS of NS matter, we have to consider two
distinct densities, $\rho={\cal E}/c^2$ and $n_{\rm b}$
(Sect.\;\ref{sect:EOS.overv}). While $\rho$ is the relevant quantity
for GTR calculations of the NS structure, it is  $n_{\rm b}$ that is
associated with an average distance between baryons (treated as
point-like objects), $r_{\rm b}\propto n_{\rm b}^{-1/3}$. Therefore,
knowing  $n_{\rm b}$,  we can compare an actual $r_{\rm b}$ with the
average distance between nucleons in nuclear matter at normal
nuclear density, $r_{_0}$, $r_{\rm b}/r_{_0}=\nbred\,^{-{1/3}}$.  At
subnuclear densities, $\rho$ of NS matter can be very well
approximated by $n_{\rm b}m_n$, where $m_n$ is neutron rest mass.
However, at supranuclear densities $\rho$ grows non-linearly with
$n_{\rm b}$. This non-linear dependence is model dependent, see
Fig.\;\ref{fig:rho-nb-eos}, and actually determines  the EOS, see Eq.
\;(\ref{eq:P.nb.rho}).
\begin{figure}[h]
\resizebox{\columnwidth}{!}
{\includegraphics[angle=0,clip]{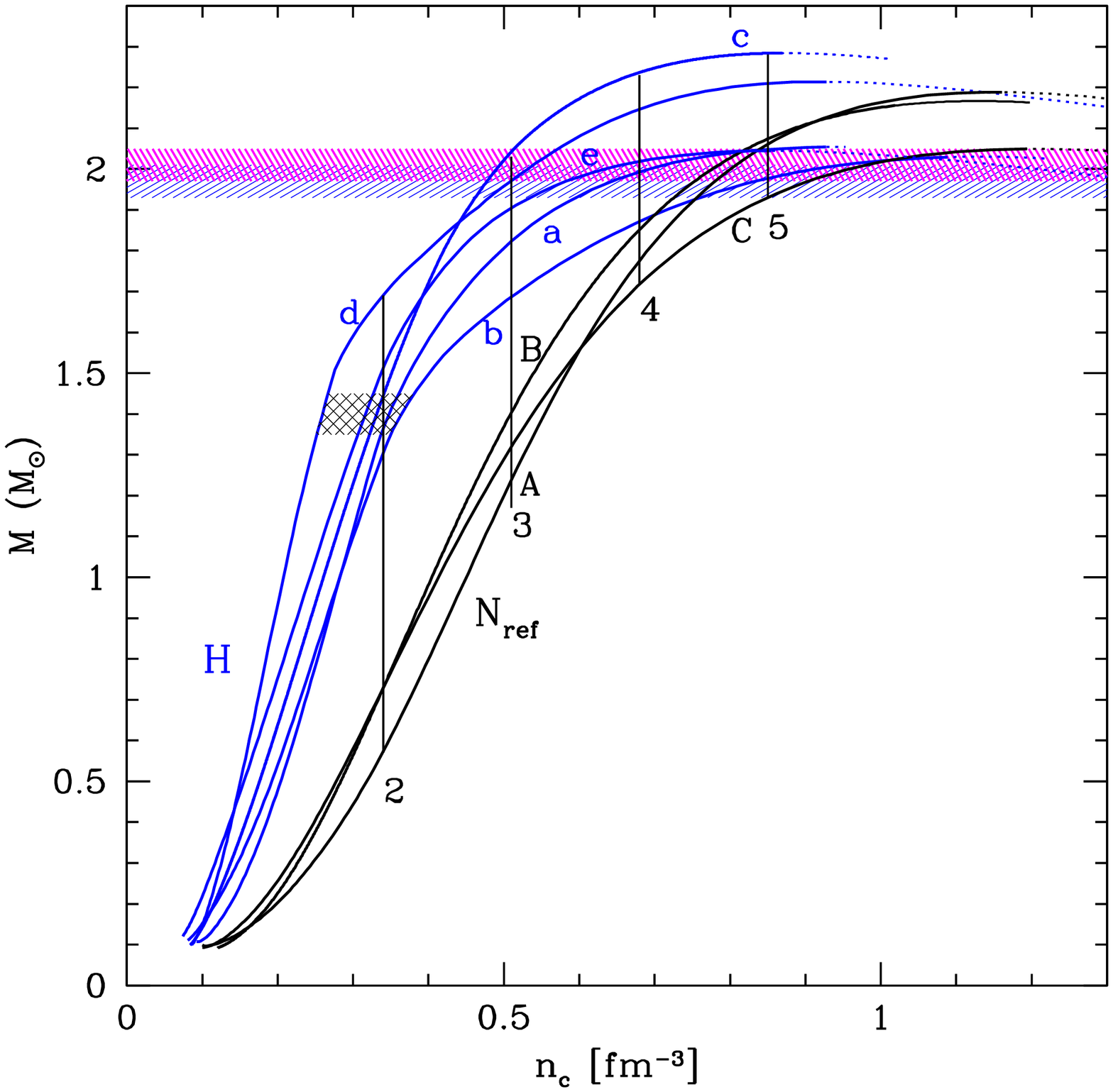}} \caption{(Colour
online) Gravitational mass $M$ vs. central baryon density $n_{\rm
c}$ for non-rotating NS  models based on the sets {\bf EOS.H} (blue
lines - H) and ${\bf EOS.N_{\rm\bf ref}}$ (black lines - ${\rm
N_{ref}}$). In ${\bf N_{\rm\bf ref}}$: {\bf A} is APR, {\bf B} is BSk20, {\bf C} is DH;
in {\bf H}: {\bf a} is SA.BSR2, {\bf b} indicates BM165, {\bf c} indicates GM1Z0, {\bf d} is UU1, {\bf e} is G.TM1C. EOS
labels from Table 1. Solid lines: stable NS configurations. Dotted
lines: configurations unstable with respect to small radial
perturbations.
 Vertical lines crossing the $M(n_{\rm c})$ curves indicate
configurations with $n_{\rm c}/\nbnuc=2,3,\ldots.$  Hatched strip
correspond to $M=1.4\pm 0.05\;\msun$, and the observational
constraints for J1614-2230 and J0348+0432 are marked in blue and
magenta, respectively (1-$\sigma$ errors).}
 \label{fig:Mnc-eos}
\end{figure}
\begin{figure}[h]
\resizebox{\columnwidth}{!} {\includegraphics[angle=0,clip]{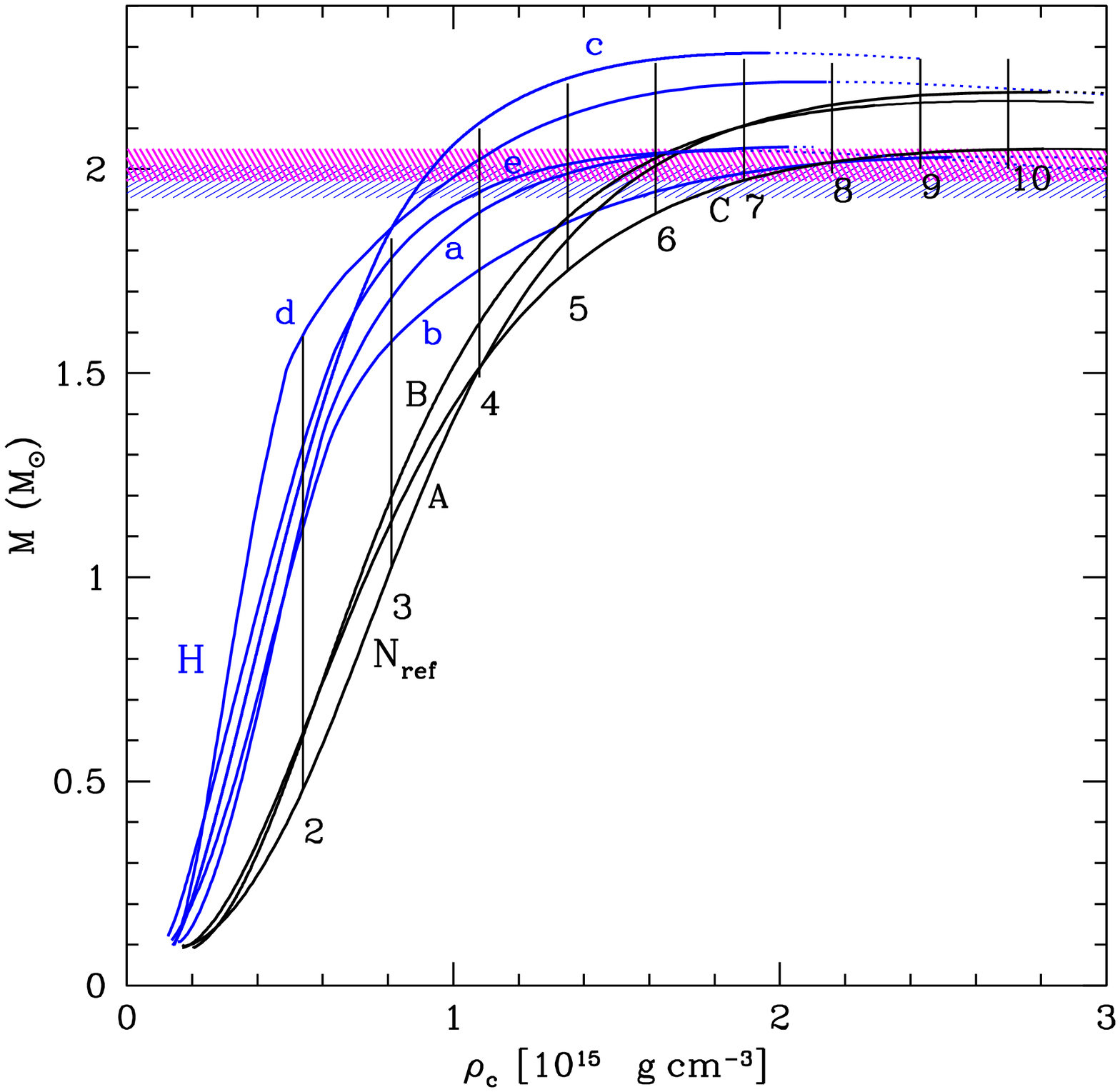}}
\caption{(Colour online) Same as in Fig.\;\ref{fig:Mnc-eos}, except with $n_{\rm c}$
replaced by $\rho_{\rm c}$.}
 \label{fig:Mrhoc-eos}
\end{figure}

In Fig.\;\ref{fig:Mnc-eos} we plot the relations between $M$ and the
central baryon density $n_{\rm c}$ for non-rotating NS models.
Several conclusions result from this figure. First, the central
density in a $2\;\msun$ NS is typically  $\nbred_{\rm c}=4 - 5$, so
that  at the star's centre $r_{\rm b}/\rnuc \approx 0.6$.  Second,
the N segment of  {\bf EOS.H,} corresponding to  $1<\nbred<2$ , is so
much stiffer than a similar segment of the ${\bf EOS.N_{\rm\bf
ref}}$, that $M^{{{\rm (H)}}}(\nbred_{\rm c}=2) \sim 2\;M^{{{\rm
(N)}}}(\nbred_{\rm c}=2)$. In other words,    to yield
$M_{\rm max}>2.0~\msun$
 despite the  hyperon softening, the pre-hyperon (nucleon) segment
 $1<\nbred<2$ of {\bf EOS.H} has to be very stiff.

Figure \ref{fig:Mrhoc-eos} shows the $M(\rhored_{\rm c})$ curves.
 The $\rhored_{\rm c}$ in $2\;\msun$ stars
can be as high as 6 - 7, significantly larger than the corresponding
values of $\nbred_{\rm c}$. For the $M_{\rm max}$ configurations, the
difference is even larger. However,  it is $n_{\rm
c}$ and not $\rho_{\rm c}$ that determines the mean inter-baryon
distance at the centre of the star.

The value of  $\nbred_{\rm c}$ for $2\;\msun$  stars can be
used to evaluate the importance of relativistic effects  in the relevant
many-body problem. The number density $n_i$ of baryon species
$i=n,p,\Lambda,\dots$ with a mass $m_i$ can be related
to their mean velocity $ \langle v_i\rangle$. In the free Fermi
gas approximation, $\langle v_i\rangle=0.26c\;(x_i\nbred)^{1/3}\left(m_{\rm n}/m_i\right)$,
where $x_i=n_i/n_{\rm b}$ and $m_{\rm n}$ is the neutron mass.
 At the centre of a $2\;\msun$ star we expect $\nbred\sim 5$ so
that $ \langle v_i\rangle \sim 0.4c\;(x_i)^{1/3}\left(m_{\rm n}/m_i\right)$. The multi-component
character of dense baryon matter implies lower $ \langle v_i\rangle$
as compared to a pure neutron matter case and, consequently, smaller
relativistic effects in the many-body system, as stressed in the
classical paper of \cite{BJ1974} to justify the use of a
non-relativistic many-body theory of dense matter. Therefore, we
might expect that not only  RMF  but also  some non-relativistic
models, consistent with semi-empirical nuclear and hypernuclear
matter constraints,  could  yield $2\;\msun$ stars with hyperonic
cores. However, as far as we know, there is only one such a
non-relativistic dense matter model satisfying these conditions
(\citealt{Yamamoto2014}).

A very recent calculation of \cite{KatSaito2014} is performed using
a relativistic formulation of G-matrix theory
(Dirac-Brueckner-Hartree-Fock approximation). Some models from this
work give  $M_{\rm max}>2\;\msun$, but they do not satisfactorily
reproduce the semi-empirical parameters of nuclear matter (see Table
I of \citealt{KatSaito2014}).

\begin{figure}[h]
\resizebox{\columnwidth}{!}
{\includegraphics[angle=0,clip]{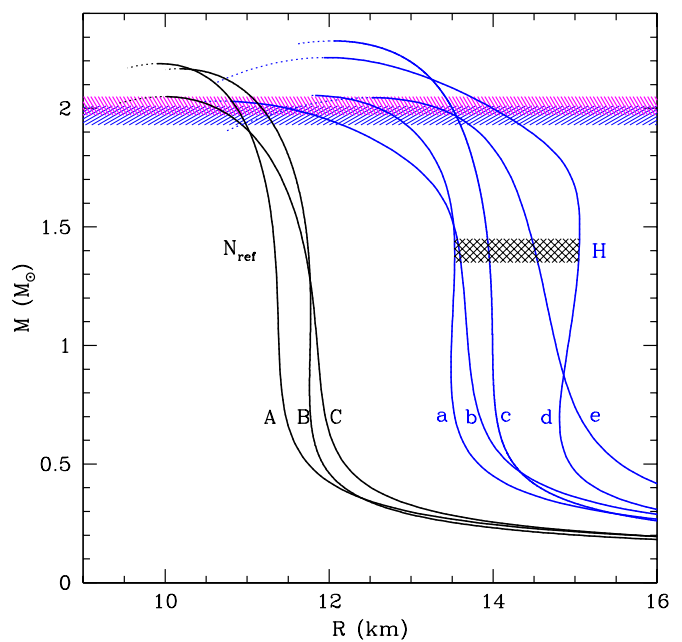}} \caption{(Colour online)
Gravitational mass $M$ vs. circumferential radius $R$ for non-rotating NS models.
For the labels, see details in the caption of figure \ref{fig:Mnc-eos}.}
 \label{fig:MR}
\end{figure}
\section{Radii of neutron stars with  hyperon cores}
\label{sect:RM.hyp}
The radius $R$  is a measurable  NS parameter  and therefore large radii of NS
with hyperon cores could be subject to an observational test. The
$M(R)$  lines for selected EOS from {\bf EOS.H} are plotted in
Fig.\;\ref{fig:MR}. By construction, selected EOS.H include
those producing an envelope of a complete H-bundle of $M(R)$ curves.

 In
the mass range $1<M/\msun<1.6$, the {\bf H}-bundle is centred
around $\sim 14.2~$km. There is a wide $>1~$km gap between the { H}
and ${\rm N_{\rm ref}}$ bundles in this mass range. More
specifically,  we find a lower bound
$R^{{(\rm H)}}>13~$km in the considered mass range.
\begin{figure}[h]
\resizebox{\columnwidth}{!}
{\includegraphics[angle=0,clip]{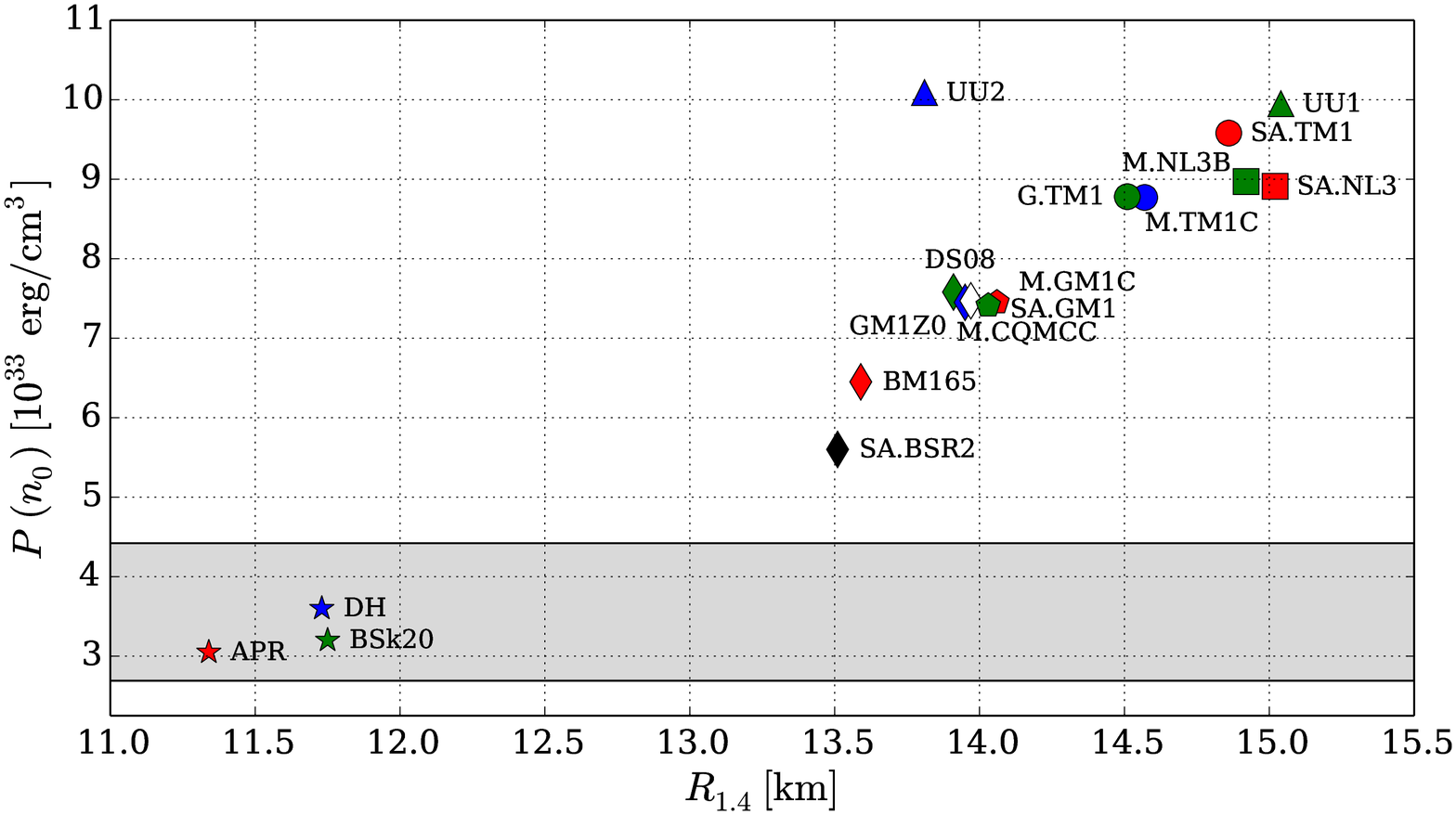}} \caption{(Colour
online) Pressure at nuclear density vs. NS radius for $1.4\;\msun$
for {\bf EOS.H} and ${\bf EOS.N_{\rm\bf ref}}$ (See table
\ref{tab:EOS} for the labels). Grey horizontal strip - range of
$P(\nbnuc)$ determined by \citet{Hebeler2013}. Circles correspond to
EOS  with the TM1, squares with NL3, pentagons with GM1 model for
the nucleon sector, respectively. Triangles indicate EOS built with
RMF models with density-dependent couplings.}
 \label{fig:pn0_r14}
\end{figure}
 In Fig.\;\ref{fig:pn0_r14} we plot the
points calculated for {\bf EOS.H} in the $P_{33}(n_{_{0}}) -
R_{1.4}$ plane, where $R_{1.4}\equiv R(1.4\;\msun)$. Large values of
$R^{\rm (H)}_{1.4}$ are correlated
\footnote{To a good approximation  $P(n_0)\propto (R^{\rm (H)}_{1.4})^4$, consistently
with \cite{LattimerPrakash2001} (D. Blaschke, private communication)}
with a high $P(\nbnuc)$
violating the upper bound of \citet{Hebeler2013}.

\begin{figure}[h]
\resizebox{\columnwidth}{!} {\includegraphics[angle=0,clip]{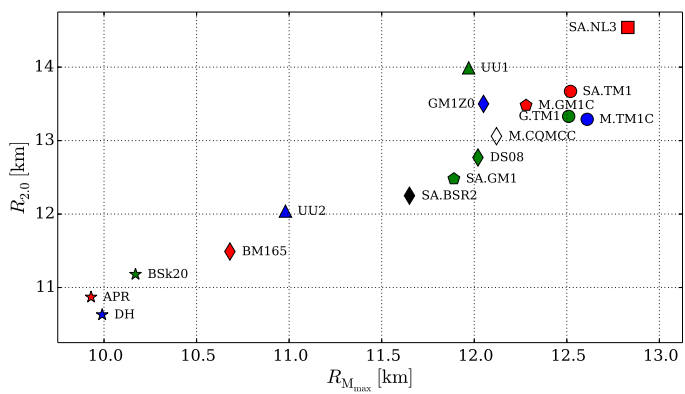}}
\caption{(Colour online) NS radius for $2.0\;\msun$ vs. radius at maximum
allowable mass. EOS labels from Table \ref{tab:EOS}.
}
 \label{fig:RmmaxR2.0}
\end{figure}
 Consider the largest, up-to-date, measured pulsar
mass, $2.0\;\msun$,  and EOS-dependent (theoretical) maximum
allowable mass $M_{\rm max}$. The EOS-dependent  radius of
$2.0\;\msun$ star and the radius at $M_{\rm max}$ are denoted by
$R_{2.0}$ and $R_{M_{\rm max}}$, respectively. Calculated  points in
the $R_{M_{\rm max}} - R_{2.0}$ plane are shown in
Fig.\;\ref{fig:RmmaxR2.0}. Those for ${\bf EOS.N_{\rm\bf ref}}$ are
tightly grouped (within less than 1 km) around $R_{2.0}= 11$ km. The
points calculated for {\bf EOS.H} are loosely distributed along the
diagonal of the bounding box, with $R^{\rm (H)}_{M_{\rm max}}$
ranging within $10.5 - 12.5$ km, and $R^{\rm (H)}_{2.0}$ within $11.5 -
14$~km.

\section{Mass vs. central pressure for neutron stars with  hyperon cores}
\label{sect:MPc}
\begin{figure}[h]
\resizebox{\columnwidth}{!} {\includegraphics[angle=0,clip]{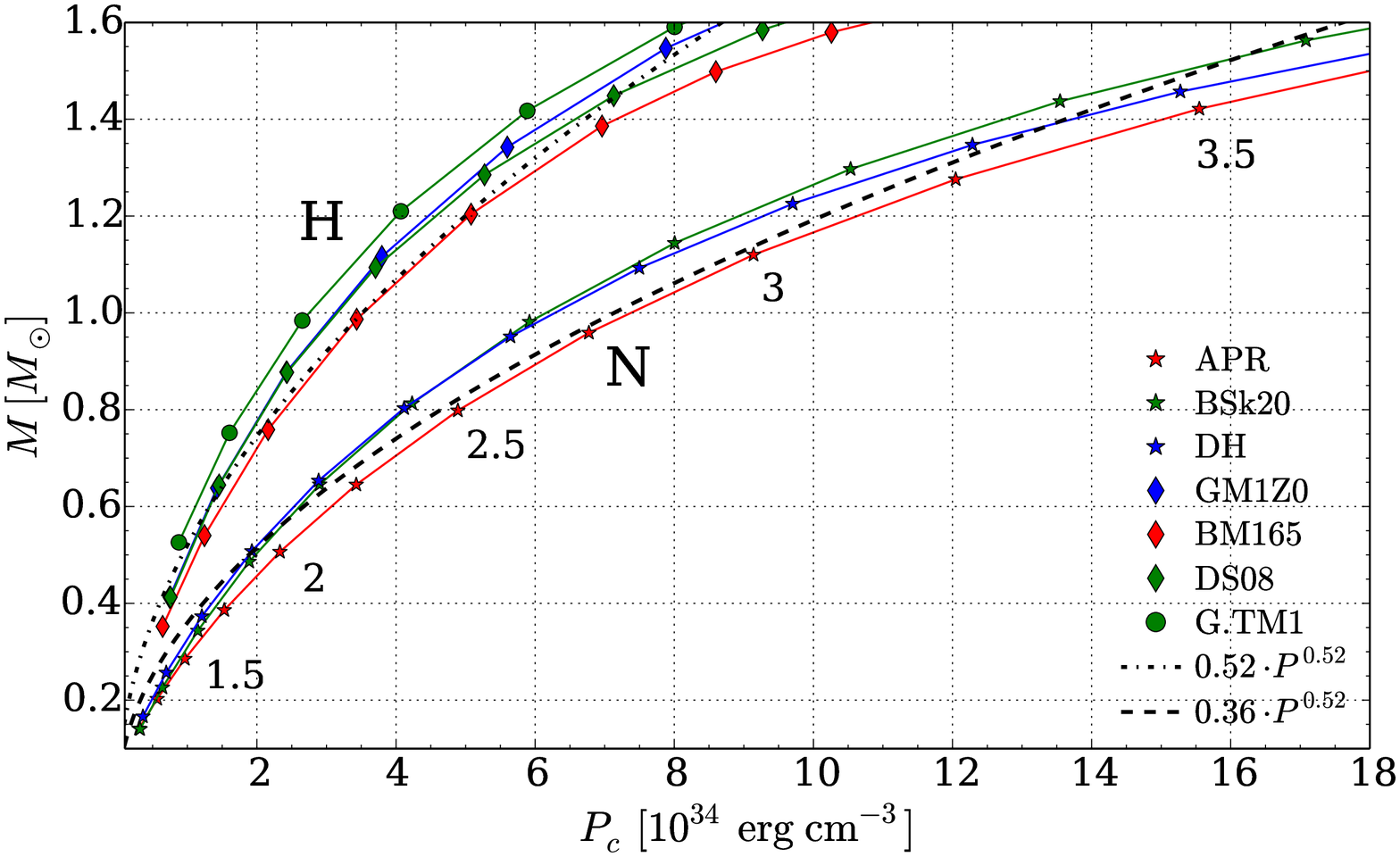}}
\caption{(Colour online)Dependence of $M$  on the central pressure, for non-rotating
NS models, for $M<1.6\;\msun$. Dashed and dashed-dotted lines describe the  $M/\msun={\cal
A}(P_{{\rm c}34})^\beta$ fits for N and H families, respectively. For details, see
the text. }
 \label{fig:MPc}
\end{figure}
The difference between  $M-R$  relations for ${\bf EOS.N_{\rm\bf
ref}}$ and {\bf EOS.H} and in particular an $R$-gap  for
$1.0<M/\msun<1.6$ reflects a difference in the pressure
distributions within NS models of a given $M$. This is visualized in
Fig.\;\ref{fig:MPc} where we show $M(P_{\rm c})$ plots for ${\bf
EOS.N_{\rm \bf ref}}$ and {\bf EOS.H} families. For $M\lesssim
1.6\;\msun$ we find that $M=M(P_{\rm c})$ is well approximated by
\begin{equation}
{M\over \msun}\simeq  {\cal A}\left(P_{\rm c,34}\right)^\beta~,
\label{eq:MPc}
\end{equation}
where  $P_{34}=P/10^{34}\;{\rm erg~cm^{-3}}$ and $\beta_{\rm
H}\approx \beta_{\rm N}=0.52$, while  prefactor ${\cal A}_{\rm
N}=0.36$  for the ${\bf N_{\rm\bf ref}}$-bundle is significantly
smaller than ${\cal A}_{\rm H}=0.52$ for the {\bf H}-bundle.

There is a sizable gap between $M^{\rm (N)}(P_{\rm c})$ and $M^{\rm
(H)}(P_{\rm c})$ bundles. The $M$-gap ranges from $\sim 0.2\;\msun$
at $P_{\rm c,34}=3$ to $\sim 0.7\;\msun$ at $P_{\rm c,34}=8$. The
fit $M\propto P_{\rm c}^\beta$ with $\beta=const.$ breaks down for
the H-family for $M>1.6\;\msun$ {because of the hyperon softening of
the EOS}. For $M\sim 2\;\msun$ (not shown in Fig.\;\ref{fig:MPc}),
the gap between the ${\bf N_{\rm\bf ref}}$ and {\bf H} bundles
disappears.

\section{On the causal bounds  on $R(M)$}
\label{sect:RM.causal}
The radius of a neutron star of given mass $M$, based on a causal EOS, cannot exceed
a limit that is calculated  by replacing this EOS above some "fiducial density"
$n_\star$  by the causal-limit (speed of sound = $c$) continuation. The
causal-limit (CL)  segment  of the EOS.CL is:
\begin{equation}
P^{\rm (CL)}=P_\star + c^2\left(\rho-\rho_\star\right)
\label{eq:CL.eos}
\end{equation}
for $n_{\rm b}\ge n_\star$, whereas $P_\star$ and $\rho_\star$ are
calculated from the original EOS, $P_\star=P(n_\star)$,
$\rho_\star=\rho(n_\star)$.

\begin{figure}[h]
\resizebox{\columnwidth}{!} {\includegraphics[angle=0,clip]{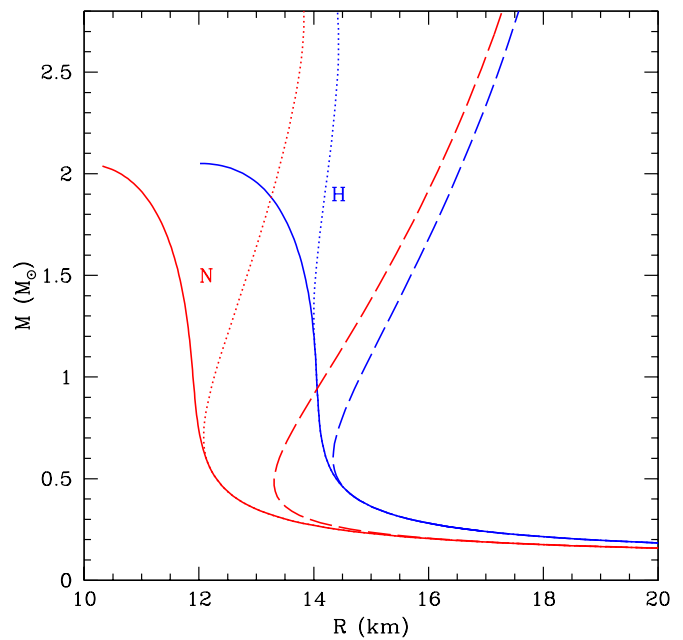}}
\caption{(Colour online) Solid lines: mass-radius relations for an EOS.N(DH) and an
EOS.H (DS08), respectively. Long-dash line: causal-limit upper bound on $R_{_M}$
for these EOS, assuming $n_\star=\nbnuc$. Dotted line: similar upper bound on
$R_{_M}$, except with $n_\star=1.8\nbnuc$, as in \citet{Hebeler2013}. }
 \label{fig:MR-CL}
\end{figure}

We choose $n_\star=\nbnuc$. Our choice is more conservative, for the reasons
 explained later in this section,   than $1.8\nbnuc$ used by \citet{Hebeler2013}.
We fix the model for the crust segment of all EOS, i.e. for
$n_{\rm b}<0.5\nbnuc$, we use the EOS described in the final fragment of
Sect.\ref{sect:EOS.B}.

\begin{figure}[h]
\resizebox{\columnwidth}{!} {\includegraphics[angle=0,clip]{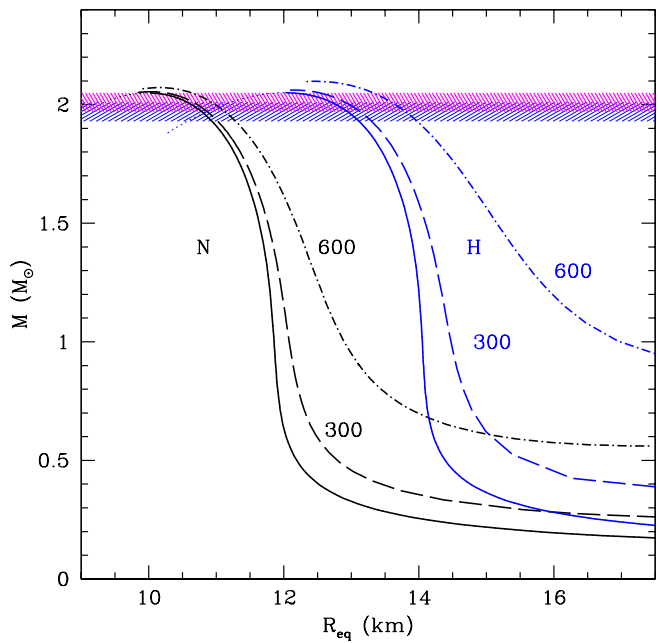}}
\caption{(Colour online) $M(R_{\rm eq})$ for two selected EOS: N (DH) and H (DS08). The 
$R_{\rm eq}$ is defined in the text. Solid curves: non-rotating configurations.
Dashed lines: configurations rotating uniformly at $f=300$ and $600$ Hz. }
 \label{fig:MRrot}
\end{figure}

In Fig.\;\ref{fig:MR-CL} we show $M(R)$ curves for {selected models
 from  ${\bf EOS.N_{\rm\bf ref}}$ and {\bf EOS.H} (DH and DS08 EOS, respectively)},
 together with their
causal-limit (CL)  bounds. The CL curves bifurcate from the actual
$M(R)$ curves at $n_{\rm c}=n_\star$, corresponding to $M=M_\star$.
For a given $M>M_\star$, the CL-bound is denoted by $R_{\rm
max}(M)$.

Consider first $\nbred_\star=1$. We get $M^{\rm (N)}_\star \approx
0.2\;\msun$ and  {$M^{\rm (H)}_\star\approx 0.5\;\msun$}. As $P^{\rm
(H)}_\star$  is significantly higher than $P^{\rm (N)}_\star$,
$R_{\rm max}^{\rm (H)}(M)$  is larger than $R_{\rm max}^{\rm
(N)}(M)$, e.g. by about 1 km at $1.4\; \msun$. The difference
decreases to 0.5 km at $2\;\msun$.

Consider now  $\nbred_\star=1.8$ chosen by \citet{Hebeler2013}. We
then get $M^{\rm (N)}_\star \approx 0.5\;\msun$ and $M^{\rm
(H)}_\star\approx 1.4\;\msun$. At $1.4\;\msun$  the actual $R^{\rm
(H)}(M)$  is larger than $R_{\rm max}^{\rm (N)}(M)\approx R_{\rm
max}^{\rm (N)}$ by  about  1.4 km. The H configuration there
violates the condition $P_{\rm c}\gg P^{\rm (H)}$ needed for a weak
dependence of $R_{\rm max}(M)$ on the EOS below $n_\star$.
Therefore, a CL-bound derived for ${\bf EOS.N_{\rm\bf ref}}$ should
not be applied to {\bf EOS.H} stars. Even at $2\;\msun$ the CL-bound for
{H} is larger by 1 km than the {\rm N}-bound. We conclude that
$\nbred_\star=1.8$ chosen by \citet{Hebeler2013} is too large to
yield a unique CL-bound for radii of the ${\bf N_{\rm\bf ref}}$ and
${\rm\bf H}$ families simultaneously.

\section{Effect of neutron star rotation}
\label{sect:RM.rot}
The fact that NSs rotate should be taken into account when comparing
their radii; the values will differ from a non-rotating solution because of
the effect of the centrifugal force.  Mature NSs, and especially
extremely stably spinning millisecond pulsars (spin frequencies larger
than 100 Hz) may be, to an excellent
approximation, described using the assumption of uniform
rotation.\footnote{At the NS birth, the temperature is high enough
to allow for differential rotation, but as soon as the star cools,
physical processes such as  convection, turbulent motions, and shear
viscosity promptly enforce uniform rotation. At sufficiently low
temperatures, the NS interior becomes superfluid. Total stellar angular
momentum is then represented by a sum of quantized vortices, but on
scales much larger than the vortex separation (which is typically much
smaller than 1 cm for known pulsars' rotation rates) fluid motion
averages out and may be considered uniform (see e.g.
\citealt{Sonin1987}).} Non-axial deformations supported by the elastic
strain accumulated in the crust, as well as non-axiality generated by the
internal magnetic field, are small for the rotating NSs that we are modelling.
Therefore, in what follows we will assume that the configurations are stationary
and axisymmetric, and we will define the circumferential equatorial radius $R_{\rm
eq}$ as the proper length of the equator divided by $2\pi$.
For a given mass $M$, the value  $R^{\rm (H)}_{\rm eq}$ for {\bf EOS.H}-based NS
increases with rotation rate more rapidly than for ${\bf
EOS.N_{\rm\bf  ref}}$. This is a direct consequence of
larger radii of hyperonic NSs. In Newtonian
terms, for fixed mass $M$ and spin frequency $f$, the gravitational
pull at the equator is weaker, while the centrifugal force is stronger
in the case of ${\rm \bf EOS.H}$ stars: both effects
increase the value of $R_{\rm eq}$. Figure\;\ref{fig:MRrot} compares
$M(R_{\rm eq})$ curves calculated for $f=300\;{\rm
Hz}$ and $f=600\;{\rm Hz}$ with the non-rotating ($f=0$) case.
Frequencies  $\sim 300$ Hz are measured for many radio and X-ray
millisecond pulsars, while $f\sim 600$ Hz is characteristic of
a group of the fastest \citep{Lorimer2008,Patruno2012}.

For the EOS in Fig.\;\ref{fig:MRrot}, the rotational increase of
$\Delta R_{\rm eq}$ is roughly twice as large for EOS.H compared with EOS.N.
It can be well approximated, for a fixed EOS and mass $M$,
as a quadratic function in $f$:
\begin{equation}
\Delta R_{\rm eq} (f,M) = R_{\rm eq} (f,M)-R (f=0,M)\simeq {\cal B}(M) f^2~.
 \label{eq:Rrot}
\end{equation}
For $M=1.4\;\msun$, one has
${\cal B}_{\rm H}=3.1\;{\rm km/kHz^2}$ and
${\cal B}_{\rm N}=1.6\;{\rm km/kHz^2}$.
At a given $f$, $\Delta R_{\rm eq}$  decreases with
increasing $M$. For an astrophysically motivated range of masses
between $1.2\;\msun$ and $1.8\; \msun$, the decrease of
$\Delta R_{\rm eq}$ with an increasing mass is approximately linear:
${\cal B}(M)$ changes
from $2.0\;{\rm km/kHz^2}$ to $1.2\;{\rm km/kHz^2}$ for EOS.N,
and from $3.8\;{\rm km/kHz^2}$ to $2.2\;{\rm km/kHz^2}$
for EOS.H. These linear approximations for ${\cal B}$ and relation (\ref{eq:Rrot})
may be used to estimate the increase of $R_{\rm eq}$ for rotating
nucleonic and hyperonic NSs.

\section{Observational determination of radii of neutron stars}
\label{sect:RM.obs}
The radius $R$ can be extracted from the analysis of
X-ray spectra emitted by the NS atmosphere. Recent attempts are based on the modelling
of X-ray emission from four different types of objects (see \citealt{Potekhin2014}
 for a review):

\begin{itemize}
\item isolated NSs (INSs);
\item quiescent X-ray transients (QXTs) in low-mass X-ray binaries (LMXBs), ie. NSs
in a binary system observed when the accretion of matter from their binary
 companion has stopped or is strongly reduced;
\item bursting NSs (BNSs) ie. NSs from which recurring and very powerful bursts,
so-called photospheric radius expansion (PRE) bursts, are observed;
\item rotation-powered radio millisecond pulsars (RP-MSPs).
\end{itemize}
The modelling of the X-ray emission, and thus the radius determination, strongly depend
on the distance to the source, its magnetic field, and the composition of its atmosphere.
 Even in the simplest case of non-rotating and non-magnetized NSs, because of
 the space-time curvature, the apparent radius $R_\infty$ that is constrained by
  the modelling actually depends on \textit{\textup{both}} the stellar radius and
   mass, $R_\infty = R/\sqrt{1-2GM/Rc^2}$.

While the magnetic field and the chemical composition of the atmosphere of INSs are unknown
and difficult to determine, QXTs have a low magnetic field, as a result of its decay due to
 the accretion of matter, and an atmosphere likely to be composed of light elements
 (H, possibly He; see \citealt{Catuneanu2013,Heinke14}). In addition, QXTs in globular clusters
  are promising sources for the mass-radius determination since their distances are well known.
BNSs that also have low magnetic fields and a light-element atmosphere are interesting sources,
all the more if the distance to them is known. However the modelling of the PRE bursts is still
subject to uncertainties and debates (see \citealt{Poutanen2014,SteinerLattBrown2013,Ozel2013}).
Finally, constraints can be derived from the modelling of the shape of the X-ray pulses observed
from RP-MSPs, in  particular, if the mass is known from radio observations \citep{Bogdanov2013}.

Conflicting constraints on the radius from the X-ray emission of QXTs, BNSs, and RP-MSPs have
been obtained by different groups. In what follows, we restrict ourselves to the
most recent publications, indicated in Table \ref{tab:Obs}. Figure \ref{fig:Obs} shows the
constraints with 2-$\sigma$ error bars derived in these papers for the radius $R_{1.4}$ of a NS
with a mass of 1.4~M$_\odot$.

Performing a simultaneous spectral analysis of six QXTs assuming unmagnetized hydrogen atmosphere, 
\citet{Guillot2014} derive  a constraint suggesting a small radius: $R=9.4\pm{1.2}$ km ($90\%$ confidence level) 
consistent with their previous results \citep{Guillot2013}.  \citet{Heinke14}, using new and archival X-ray data, 
re-performed an analysis of the spectrum
of the one of these QXTs, NGC 6397 and argue that a helium atmosphere is favoured over a hydrogen atmosphere such as
the one used in \citet{Guillot2013}, which favours a larger radius.

For all papers in Table \ref{tab:Obs}, if not available, constraints on $R_{1.4}$ with a
 2-$\sigma$ error bars were derived assuming a Gaussian distribution for the radius,
 which makes the comparison between the results easier. The
constraint from \citet{Ozel2013}
 is extracted from their figure 3.
\begin{table*}
\begin{center}
\caption{Radius determination: most recent publications. For each paper, first column:
abbreviation used in figure \ref{fig:Obs}, second column: objects and type of objects used,
 third column: reference. (a): \citet{Poutanen2014}; (b):
\citet{Guillot2014}; (c): \citet{Ozel2013};
  (d): \citet{Heinke14}; (e): \citet{SteinerLattBrown2013};  (f): \citet{Bogdanov2013}; (g):
  \citet{Verbiest2008}. See text for details.}
\begin{tabular}{clc}
\hline
\hline
Abbreviations & Objects & References \\
\hline
\hline
PN14 & BNS: 4U 1608-522 & (a) \\
GR14 & QXT: M28, NGC 6397, M13, $\omega$ Cen, NGC 6304, M30  & (b)\\
GO13 & BNS: SAX J1748.9-2021 & (c,d)     \\
SL13 & BNS: 4U 1608-522, KS 1731-260, EXO 1745-248, 4U 1820-30 & (e) \\
     & QXT: M13, 47 Tuc, NGC 6397, $\omega$ Cen &   \\
B13  & RP-MSP: PSR J0437-4715 & (f,g)\\
\hline
\hline
\end{tabular}
\label{tab:Obs}
\end{center}
\end{table*}

\begin{figure}[h]
\resizebox{\columnwidth}{!} {\includegraphics[angle=-180,clip]{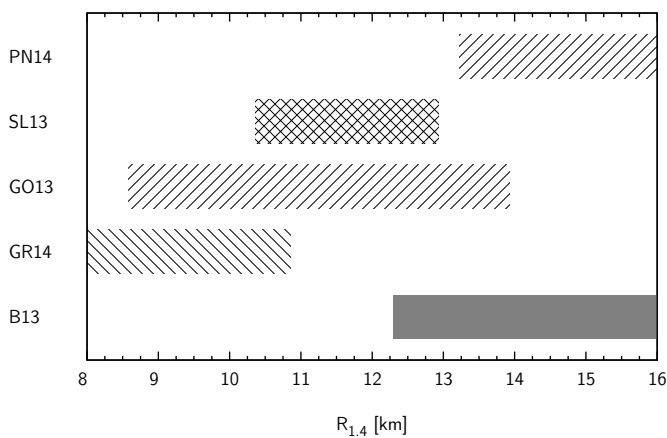}}
\caption{Observational constraints on the radius $R_{1.4}$ of a 1.4~M$_\odot$ NS from the most recent publications. See Table \ref{tab:Obs} for the labels and text for details. The 2-$\sigma$ error-bars are plotted.}
 \label{fig:Obs}
\end{figure}

Formally,  at 2-$\sigma$ level,  all constraints on $R_{1.4}$ but those by \citet{Poutanen2014} and \cite{Guillot2014} are
consistent with one another and give the radius of $\sim12.8\;$km  for $M=1.4\;\msun$.
All present EOS.H have a substantially larger radius $R_{1.4}\geq 13.51$ km as indicated in
Table \ref{tab:P0.R.RCL} and Fig. \ref{fig:pn0_r14} and thus violate this upper bound. They are
 consequently consistent only with the {\it lower bounds} on $R_{1.4}$
obtained by \citet{Poutanen2013} and \citet{Bogdanov2013}.

However, the determination of the radius of NSs is subject to many assumptions,
uncertainties, and systematic effects, (see e.g. Table 1 in \citealt{Potekhin2014}).
In addition, the inclusion of rotation strongly complicates the analysis of the collected X-ray
spectra from both QXTs and BNSs, which are likely to rotate at a frequency of few hundreds of Hz. Rotation
is expected to affect the radius determination by $\sim10\%$ according
to \citet{Poutanen2014} and \citet{BO14}.

\section{Summary, discussion,  and conclusions}
\label{sect:conclusions}
The observational constraint that $M_{\rm max}>2.0\;\msun$ for NS with
hyperon cores imposes a fine tuning of the dense matter model, which
 has to be simultaneously consistent    with semi-empirical nuclear
and hyper-nuclear parameters. With one very recent exception
\citep{Yamamoto2014}, which will be discussed separately, only
specific types of non-linear RMF models are able to satisfy this
constraint. The EOS of NS matter for these models, forming the set
{\bf EOS.H},  have the  features that are described below.
\vskip 2mm
\parindent 0pt

{\it Overpressure at  $\nbnuc$}. Pressures $P^{\rm (H)}(\nbnuc)$ are
significantly  higher than the robust upper bound obtained by
\citet{Hebeler2013}. Introducing a density-dependence for the coupling
constants in the RMF Lagrangian does not help in this respect.

\vskip 2mm {\it Large radii}. For $1.0<M/\msun<1.6,$  one obtains $13<R^{\rm(H)}/{\rm km}<15$. These radii are consistent at 2-$\sigma$
confidence level only  with two lower bounds out of the five most
recent constraints, derived by analyzing and modelling the X-ray
emission from NSs in quiescent LMXBs, from those exhibiting
photospheric expansion bursts, and from radio millisecond pulsars.
 Future simultaneous determinations
 of $M$ and $R$ through analysis of the X-ray spectra with a $\sim5\%$ precision, thanks to
 the forthcoming  NICER \citep{NICER}, Athena+ \citep{Athena+} and possibly
 LOFT \citep{LOFT} missions, could either rule out hyperon cores
 in NS or leave open the possibility of sizable hyperon cores.

\parindent 21pt

Overpressure at $\nbnuc$ and  large radii are likely to be
interrelated. To get $M_{\rm max}>2\;\msun$, an EOS.H should
necessarily have unusually stiff pre-hyperon segment, and this
results in large radii for $M\lesssim 1.6\;\msun$. However, in our
opinion the {\it \textup{upper bound}} on $P(\nbnuc)$ by \citet{Hebeler2013} is
sufficiently robust to be respected. Therefore, we propose
including this constraint in the procedure of determination of the RMF
Lagrangian coupling constants, in addition to the standard fitting
four parameters $E_{\rm s},n_{\rm s},E_{\rm sym}, K_{\rm s}$. In the
weakest form, this constraint would read
\begin{equation}
{P^{\rm (PNM)}(\nbnuc)\leq P_{\rm
max}^{\rm(Heb)}(\nbnuc)=5.4\times10^{33} \rm{\ dyn\ cm}^{-2}},
\label{eq:PNM.Heb}
\end{equation}
 where $P^{\rm (PNM)}$ is the pressure of pure neutron matter (PNM)
 calculated for a given RMF model, and $P_{\rm max}^{\rm({Heb})}$ is the
upper bound to this pressure obtained by \citet{Hebeler2013}.
We prefer to use constraint (\ref{eq:PNM.Heb}) instead of fitting the very uncertain
 $L_{\rm s}$  , referring to weakly asymmetric nuclear matter.
We are now studying the possibility of imposing    both the PNM constraint (\ref{eq:PNM.Heb}) and getting
$M_{\rm max}>2\;\msun$ to  narrow the {\bf EOS.H} family of  RMF models.

A very recent model of EOS.H fulfilling $M_{\rm max}>2\;\msun$
\citep{Yamamoto2014} deserves a separate discussion. This model does
not rely on the RMF approximation, rather it is\  calculated using the
G-matrix theory. A crucial new element is a strong three- and
four-baryon repulsion resulting from the multi-pomeron exchanges
between baryons. The many-body theory is applied to a number of
terrestrial nuclear and hyper-nuclear data, which are sufficient to
fit  three sets of parameters of the models. The set MPa  yields the
stiffest hyperon NS-cores  and is the only set satisfying  $M_{\rm
max}>2\;\msun$. The MPa curves in figures 8,\,9 and 11 of
\citet{Yamamoto2014} indicate that the MPa EOS has similar basic
features as those characteristic of the RMF models of our
{\bf EOS.H} set. Namely,  $R^{\rm (MPa)} \gtrsim 13.5~$km for
$1.2<M/\msun<1.6$ and $M^{\rm (MPa)}(\overline{n}_{\rm c}=2)\approx
1.3~\msun$, which is twice the value characteristic of our ${\bf
EOS.N_{\rm\bf ref}}$ set. This indicates that the pre-hyperon
segment of the MPa EOS is very stiff , in particular, with a large overpressure
at $\nbnuc$. In conclusion, these
features of the MPa EOS  coincide with those of our RMF-{\bf
EOS.H} set, and should be subject to the same tests.

As far as observations are concerned,  owing to the current
large uncertainties on the radius determination that exist because of
assumptions in the models and systematic effects, no stringent
conclusion on the radius of a $1.4\;\msun$ NS can be derived.

In our opinion, a robust observational upper bound on $R$ will
become available only with advent  of high-precision X-ray
astronomy, like that promised by the NICER, Athena+, and LOFT
projects. A simultaneous measurement of $M$ and $R$ within a
few percent error is expected to be achieved, and then  used in
combination with a maximum measured pulsar mass (at present $2.01\pm
0.04\;\msun$) as a robust criterion in our quest to unveil the
structure of neutron star cores.

\begin{acknowledgements}
 We thank V. Dexheimer, M. Oertel, A. Sulaksono, H. Uechi, and Y. Yamamoto  for providing
 us with the EOS tables. We are grateful to H. Uechi and Y. Yamamoto for helpful comments
 concerning their EOS. Correspondence with N. Chamel and A. Fantina about the BSk EOS was very helpful.
 We are grateful to J.M. Lattimer for his comments after a talk by one of the authors
 (PH) during the EMMI meeting a FIAS (Frankfurt, Germany, November
 2013). We thank M. Oertel and M. Hempel for useful discussions. We are also grateful to
 D. Blaschke for his helpful remarks on the meaning of Fig. \ref{fig:pn0_r14}.
 One of the authors (MF) was supported by the  French-Polish LIA  HECOLS  and by
 the Polish NCN HARMONIA grant DEC-2013/08/M/ST9/00664.
 This work was partially supported by the Polish NCN grant no  2011/01/B/ST9/04838.
\end{acknowledgements}


\appendix
\section{Collected numerical results  for {\bf EOS.N} and {\bf EOS.H}}
\label{sect:Appendix}
\begin{table*}
\begin{center}
\caption{Parameters of the EOS and of  NS models based on them.}
\begin{tabular}[t]{cccccccc}
\hline\hline
 EOS & $P(n_{_0})$ & $\rho(n_{_0})$ & $R^{\rm (CL)}_{1.4}$ &
$R_{1.4}$  & $L_{\rm s}$ &  $R_{M_{\rm max}}$ & $M_{\rm max}$ \\
  &  $(10^{33}~{\rm {dyn}\over {cm^2}})$ & $({\rm {10^{14}g\over {cm^3}}})$ &
  (km)  &  (km) & (MeV) &  (km) & $\msun$ \\
\hline\hline
APR     & 3.05  & 2.72 & 15.01    & 11.34 & 59  & 9.93  & 2.19\\
BSk20   & 3.20  & 2.72 & 14.95    & 11.75 & 37  & 10.18 & 2.17\\
DH      & 3.60  & 2.72 & 15.03    & 11.73 & 46  & 9.99  & 2.05\\
\hline\hline
BM165   & 6.45  & 2.74 & 15.46    & 13.59 & 74  & 10.68 & 2.03\\
DS08    & 7.58  & 2.74 & 15.52    & 13.91 & 88  & 12.02 & 2.05\\
GM1Z0   & 7.45  & 2.72 & 15.51    & 13.95 & 94  & 12.05 & 2.29\\
M.CQMCC & 7.47  & 2.73 & 15.61    & 13.97 & 91  & 12.12 & 2.08\\
SA.BSR2 & 5.60  & 2.70 & 15.40    & 13.51 & 62  & 11.65 & 2.03\\
SA.TM1  & 9.58  & 2.82 & 16.35    & 14.86 & 110 & 12.52 & 2.10\\
G.TM1   & 8.78  & 2.75 & 15.91    & 14.51 & 110 & 12.51 & 2.06\\
M.TM1C  & 8.77  & 2.74 & 15.94    & 14.57 & 111 & 12.61 & 2.03\\
SA.NL3  & 8.91  & 2.72 & 16.14    & 15.02 & 118 & 12.83 & 2.32\\
M.NL3B  & 8.97  & 2.74 & 15.98    & 14.92 & 118 & 13.18 & 2.07\\
M.GM1C  & 7.45  & 2.72 & 15.61    & 14.06 & 94  & 12.28 & 2.14\\
SA.GM1  & 7.41  & 2.71 & 15.64    & 14.03 & 94  & 11.98 & 2.02\\
UU1     & 9.95  & 2.72 & 15.78    & 15.04 & 117 & 11.97 & 2.21\\
UU2     & 10.09 & 2.73 & 15.79    & 13.81 & 117 & 10.98 & 2.12\\
\hline\hline
\end{tabular}
\end{center}
\label{tab:P0.R.RCL}
\end{table*}


\end{document}